\documentclass{jfm}

\usepackage{graphicx}
\usepackage{natbib}

\ifCUPmtlplainloaded \else
  \checkfont{eurm10}
  \iffontfound
    \IfFileExists{upmath.sty}
      {\typeout{^^JFound AMS Euler Roman fonts on the system,
                   using the 'upmath' package.^^J}%
       \usepackage{upmath}}
      {\typeout{^^JFound AMS Euler Roman fonts on the system, but you
                   dont seem to have the}%
       \typeout{'upmath' package installed. JFM.cls can take advantage
                 of these fonts,^^Jif you use 'upmath' package.^^J}%
      }
  \else
  \fi
\fi

\ifCUPmtlplainloaded \else
  \checkfont{msam10}
  \iffontfound
    \IfFileExists{amssymb.sty}
      {\typeout{^^JFound AMS Symbol fonts on the system, using the
                'amssymb' package.^^J}%
       \usepackage{amssymb}%
         \let\leq=\leqslant
         
      }{}
  \fi
\fi

\ifCUPmtlplainloaded \else
  \IfFileExists{amsbsy.sty}
    {\typeout{^^JFound the 'amsbsy' package on the system, using it.^^J}%
     \usepackage{amsbsy}}
    {}
\fi

\usepackage{caption}
\usepackage{subcaption}

\newcommand\Rey{\mbox{\textit{Re}}}  
\newcommand\Bi{\mbox{\textit{Bi}}}            

\newsavebox{\astrutbox}
\sbox{\astrutbox}{\rule[-5pt]{0pt}{20pt}}

\title[Bifurcation in a thin film flowing over a locally heated surface]{Bifurcation in a thin liquid film flowing over a locally heated surface}

\author[H. H. Katkar and J. M. Davis]%
{Harshwardhan H. Katkar%
,\ns
Jeffrey M. Davis}

\affiliation{Department of Chemical Engineering, University of Massachusetts,
686 N Pleasant St, MA 01003, USA}

\pubyear{2010}
\volume{650}
\pagerange{119--126}
\date{}
\begin{document}

\maketitle

\begin{abstract}
We investigate the non-linear dynamics of a two-dimensional film flowing down a finite heater, for a non-volatile and a volatile liquid. An oscillatory instability is predicted beyond a critical value of Marangoni number using linear stability theory. Continuation along the Marangoni number using non-linear evolution equation is used to trace bifurcation diagram associated with the rivulet instability. Hysteresis, a characteristic attribute of a sub-critical Hopf bifurcation, is observed in a critical parametric region. The bifurcation is universally observed for both, a non-volatile film and a volatile film.
\end{abstract}


\section{Introduction}

Thin liquid films flowing over heated surfaces are commonly encountered in the industry. Examples of industrial processes involving such systems include falling film evaporators, spin coating, etc. Temperature variations along the interface cause surface tension gradients that result into Marangoni stresses, typically leading to formation of patterns triggered by instabilities.  An example is the formation of regular hexagonal pattern, also known as Benard cells, in a thin liquid film resting on a uniformly heated horizontal substrate. If heating is non-uniform, yet more interesting patterns are observed.

Experimental work reported in literature on non-volatile liquid films flowing on a heater under gravity show a rich variety of such patterns.  Above a certain heat flux from the heated surface, a horizontal roller (capillary ridge) is formed near the leading edge of the heater as a result of Marangoni stress \citep{Kabov1998}. This results into formation of transverse thermal gradients, causing the flow to break into an array of vertical rivulets with a predominant wavelength. \citet{Kabov2002} further investigated the formation of these rivulets by measuring the temperature distribution and found that temperature gradients exist in both stream-wise and span-wise directions. Stagnation zones are observed near the capillary ridge in tracer particle experiments, with the particles moving slowly in the span-wise direction until they reach a rivulet streamline. \citet{Frank2003} observed a similar phenomenon in three dimensional direct numerical simulations of a film flowing over a heater, modeled with a step-wise temperature distribution with parameters fitted to match experimental film profiles.  A strong reverse flow is observed to be developed near the capillary ridge, which breaks up into rivulets as soon as the film experiences a small disturbance in span-wise direction. \citet{Frank2006} investigated flow of thin films of a non-volatile liquid falling down a vertical substrate with an embedded heater experimentally and numerically by solving three dimensional Navier-Stokes equations for the film.  Parametric study revealed the existence of a critical value for the Marangoni number ($M$) for onset of rivulet instability.  The critical Marangoni number was found to increase non-linearly with Reynolds number (\Rey).

On the theoretical front, a number of different approaches have been proposed in the literature to understand the dynamics of these systems. \citet{Kalliadasis2003} modeled a thin non-volatile film falling down a local heater using an integral boundary layer approximation (long wave theory), which reduces to lubrication approximation with additional condition of negligible inertia.  With the boundary conditions of uniform thickness and ambient temperature far from the heater, the steady state solution obtained from such models predict a film with a capillary ridge near the leading edge of the heater and a depression downstream.  Linear stability analysis for span-wise disturbances reveals both continuous and discrete spectra of eigenvalues.  The discrete spectrum is found to govern stability of the film, with the film being unstable above a critical Marangoni number (rivulet instability). Analysis based on a similar long wave approximation with negligible inertia for an array of heaters shows a varied behavior depending on the value of Biot number ($Bi$) \citep{Skotheim2003}.  For $Bi=0$, linear stability analysis shows that while there exists a range of unstable modes for perturbations in the span-wise direction that describe rivulet instability, the flow is stable for perturbations in the stream-wise direction. On the other hand, for $Bi \neq 0$, unstable modes are observed for perturbations even in the stream-wise direction. The film oscillates in the stream-wise direction as a result of this oscillatory instability.  Such an oscillatory instability is also observed in films flowing down a uniformly heated substrate.

For a volatile film, additional effects due to evaporation or condensation of the liquid have to be considered in the model.  \citet{Burelbach1988} modeled a horizontal static volatile liquid on a uniformly heated substrate using a one sided evaporation model, assuming that the density, viscosity and thermal conductivity of vapor are negligible compared to those of the liquid.  In an isothermal film with no thermocapillarity, disturbances of a certain range of wavelengths grow in time to cause film rupture.  On the other hand, an evaporating film with a dynamic base state gives a time dependent range of unstable modes. The growth rate of disturbances increases with thinning of the film due to vapor-recoil instability caused by the normal force exerted by the vapor on liquid near the interface. In case of non-equilibrium evaporation, where the interfacial temperature is not constant across the interface, growth rate of instabilities depends on balance between the stabilization due to non-equilibrium and destabilization due to thermocapillarity causing a change in rupture time. The dynamics of a volatile film are thus affected by the volatility of the liquid among other parameters.

Evolution equation for film flowing over a locally heated substrate was developed using a lubrication approximation in a three dimensional flow by \citet{Tiwari2007}. They used finite length heater model similar to the one discussed in the next section. Linear stability analysis revealed existence of stable continuous spectrum with bounded oscillations far downstream and discrete spectrum with localized modes, similar to \citet{Skotheim2003}.  A band of discrete modes not including zero wavenumber is found to be unstable above a critical Marangoni number, resulting into rivulet instability. This model based on lubrication approximation was further modified by \citet{Tiwari2009} to include the effect of evaporation.  It was shown that the equation for interfacial temperature considering a one-sided evaporation model \citep{Burelbach1988} is functionally similar to that derived from a vapor mass transfer limited evaporation model.

Two different types of instabilities were reported for a volatile liquid.  At one critical value of Marangoni number, a rivulet instability was observed. Additionally, another critical Marangoni number characterized by a band of complex eigenvalues with positive real parts was found, above which the film exhibits a thermocapillary oscillatory instability.  Linear stability analysis predicted that evaporation is destabilizing for both instabilities. The leading eigen functions were found to be more pronounced over the heater, where the film has highest interfacial temperature. When disturbances in span-wise directions are negligible, these eigen functions are oscillatory in nature, resulting from the competition between high Marangoni stresses developed due to large evaporative fluxes and gravity.  The oscillations were found to decay far from the heater.  Energy analysis predicted the primary cause for observing oscillatory instability to be thermocapillary flow in the stream-wise direction, due to film perturbations that result into significant variations in interfacial temperature, further enhanced with large evaporative flux at the interface.

The aim of this work is to systematically investigate the oscillatory instability observed by \citet{Tiwari2009} for a finite length heater, over a range of parameter space. Since stream-wise disturbances are responsible for the oscillatory instability, a two-dimensional thin liquid film is used to study the effect of stream-wise disturbances on film stability. Two types of liquids are considered- a non-volatile liquid and a volatile liquid. Linear stability analysis is performed for non-volatile films showing the dependence of Marangoni number and Biot number on the onset of oscillatory instability. System dynamics are studied by integrating the full non-linear evolution equation. A continuation along Marangoni number captures the onset of oscillatory instability in agreement to the predictions from linear stability analysis. Further, a hysteresis region is found to exists within a range of Marangoni numbers, which is identified as a result of a sub-critical Hopf bifurcation. Similar computations are also performed for a volatile film for which an additional evaporation term appears in the evolution equation.  The existence of sub-critical Hopf bifurcation is found to be universally present for both types of films studied, within the range of parameters studied.


\section{Problem Formulation}\label{sec:problem_formulation}
\begin{figure}
\begin{center}
\includegraphics[width=0.48\textwidth]{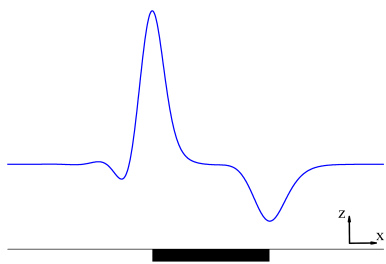}
\caption{Schematic of the system}
\label{fig:schematic}
\end{center}
\end{figure}
Consider a two dimensional thin liquid film flowing on a locally heated surface, as shown in figure \ref{fig:schematic}. The coordinate system is chosen such that flow is along the direction of positive $x$ axis and $z$ is normal to the surface. The heater surface has a temperature distribution $T_0(x)$ which results into non-uniform temperature distribution inside the film, forming a capillary ridge near the leading edge of the heater.

\citet{Tiwari2007} use lubrication approximation, no slip boundary conditions along with known temperature distribution $T_0(x)$ at the base and stress balance and heat flux balance along the air-film interface to derive the evolution equation for film thickness $h(x,t)$ as
\begin{equation}
\label{lh_evolutioneqn_dim}
\frac{\partial h}{\partial t} + \frac{1}{\mu}\frac{\partial}{\partial x}\left[\frac{h^3}{3} \left(\rho g + \gamma_{0} \frac{\partial^3 h}{\partial x^3} \right) - \frac{\gamma_{T} h^2}{2} \frac{\partial T^i}{\partial x} \right] = 0
\end{equation}
where $g$ is the acceleration due to gravity and $\rho, \mu$ are the liquid density and viscosity respectively. 
The surface tension $\gamma$ is assumed to linearly vary with temperature $T$, as $\gamma(T) = \gamma_{0} - \gamma_{T} (T-T_{\infty})$. $\gamma_{0}$ is the surface tension at a reference temperature $T_{\infty}$ far from the heater and $\gamma_T > 0$ is the absolute rate of change of surface tension with temperature.

$T^i$ is the temperature at the interface ($z=h$) obtained from solving the energy balance, given by $T^i-T_{\infty} = (T_0 - T_{\infty})/[1+\frac{h_{t}}{k_{t}} h]$ where $h_{t}$ and $k_{t}$ are the heat transfer coefficient of air and thermal conductivity of the liquid respectively.

The film thickness upstream, far from the heater, $h_{\infty}$ is used as a length scale along the $z$ coordinate. A dynamic length scale $l_c \equiv h_{\infty}(3Ca)^{-1/3}$  is used as an appropriate length scale in $x$ direction. Here, $Ca \equiv \mu u_c/\gamma_0$ is the Capillary number.  The balance of viscous and gravity terms gives rise to a velocity scale in the flow direction, $u_c = \rho g h_{\infty}^2/(3 \mu)$.  The dimensionless coordinates are thus defined as $X = x/l_c$, $H = h/h_{\infty}$. Substituting these in the evolution equation and using a dimensionless temperature $\theta = (T - T_{\infty})/\Delta T$, where $\Delta T$ is the temperature increase at the heater, the dimensionless form governing the evolution of the film is obtained as
\begin{equation}
\label{lh_evolutioneqn}
\frac{\partial H}{\partial t} + \frac{\partial}{\partial X}\left[{H^3} \left(1 + \frac{\partial^3 H}{\partial X^3} \right) - M H^2 \frac{\partial \theta^i}{\partial X} \right] = 0
\end{equation}
The above equation has two parameters. The number $M \equiv (\epsilon \gamma_T \Delta T)/(2 \mu u_c)$ is the dimensionless Marangoni number, with $\epsilon = h_{\infty}/l_c \ll 1$. Dimensionless temperature distribution at the interface is given by
\begin{equation}
\label{lh_thetai}
\theta^i(X) = \frac{\theta_0(X)}{[1+Bi\, H]}
\end{equation}
where $Bi \equiv h_{t} h_{\infty} /k_{t}$ is the dimensionless Biot number that characterizes the ratio of convective heat transfer to the air to conduction in the liquid.

For a volatile film, an additional term appears in the evolution equation resulting from the loss of mass by evaporation. \citet{Tiwari2009} discuss various limits for the evaporation term, and derive a generic expression for evolution of the liquid film as
\begin{equation}
\label{lh_evolutioneqn_volatile}
\frac{\partial H}{\partial t} + E J + \frac{\partial}{\partial X}\left[{H^3} \left(1 + \frac{\partial^3 H}{\partial X^3} \right) - M H^2 \frac{\partial \theta^i}{\partial X} \right] = 0
\end{equation}
where $E = (k_{t}\Delta T)/(\rho u_c h_{\infty} \Delta H_{vap})$, dimensionless evaporation number, characterizes the ratio of time scale of convection to that of evaporation. Here $\Delta H_{vap}$ is the latent heat of vaporization. The dimensionless mass flux is given by $J = \theta_0/(K+H)$, with $K=Bi^{-1}$ as a parameter. In the limit of $E \to 0$, equation \ref{lh_evolutioneqn_volatile} reduces to the evolution equation for non-volatile film given by equation \ref{lh_evolutioneqn}.

For a non-volatile film, the appropriate boundary conditions are those of a flat film far from the heater.
\begin{equation}
\begin{array}{c c c c}
\label{lh_evolutioneqn_bcs}
H \to 1,& H_{XXX} \to 0 & as & X \to \pm \infty
\end{array}
\end{equation}
In practice, at steady state, the film is uniform at a distance of $X = \pm 80$ from the center of the heater.
For a volatile film, mass is lost due to evaporation. Hence, film thickness far from the heater downstream will be different than that upstream. In order to facilitate this, the downstream boundary conditions in equation \ref{lh_evolutioneqn_bcs} are replaced by
\begin{equation}
\begin{array}{c c c c}
H_X \to 0, & H_{XXX} \to 0 & as & X \to \infty
\label{vf_evolutioneqn_bcs}
\end{array}
\end{equation}
These boundary conditions are used for both non-volatile and volatile films for direct comparison. In equation \ref{lh_thetai}, a known temperature $\theta_0(X)$ is assumed at the base. To approximate a finite length heater the following temperature profile is used instead of a Gaussian distribution.
\begin{equation}
\label{lh_To}
\theta_{0} (X) = 0.5 \{tanh [ \omega (X + \chi) ] - tanh[0.5(X-8)]\}
\end{equation}
with the parameters $\omega = 1$ and $\chi = 4$.

\section{Results}\label{sec:results}
\subsection*{Non-volatile film}
\begin{figure}
\centering
	\begin{minipage}[t]{0.48\textwidth}
		\centering
		\includegraphics[width=\textwidth]{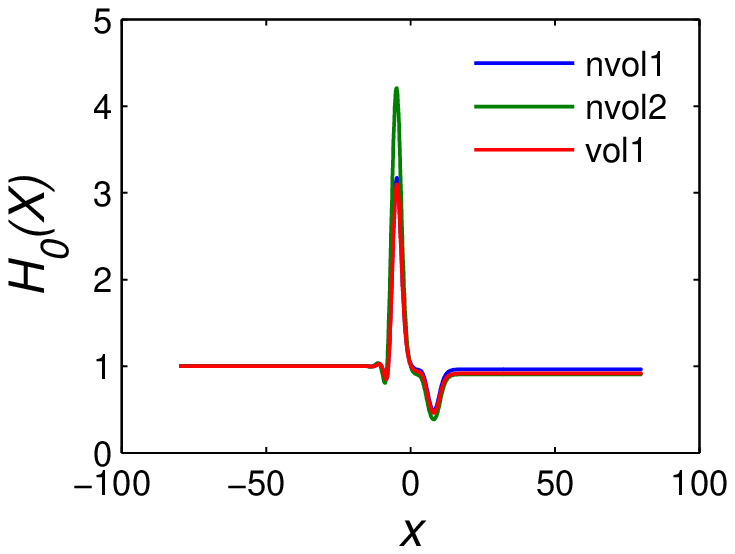}
		\captionof{figure}{$H_0(X)$ profiles at different parameter values. $nvol1 \equiv (Bi=0.125,M=14)$, $nvol2 \equiv (Bi=0.07,M=19)$ and $vol1 \equiv (K=8, E = 10^{-1},M=13.5)$.}
		\label{fig:baseprofiles}
	\end{minipage}
	\quad
	\begin{minipage}[t]{0.48\textwidth}
		\centering
		\includegraphics[width=\textwidth]{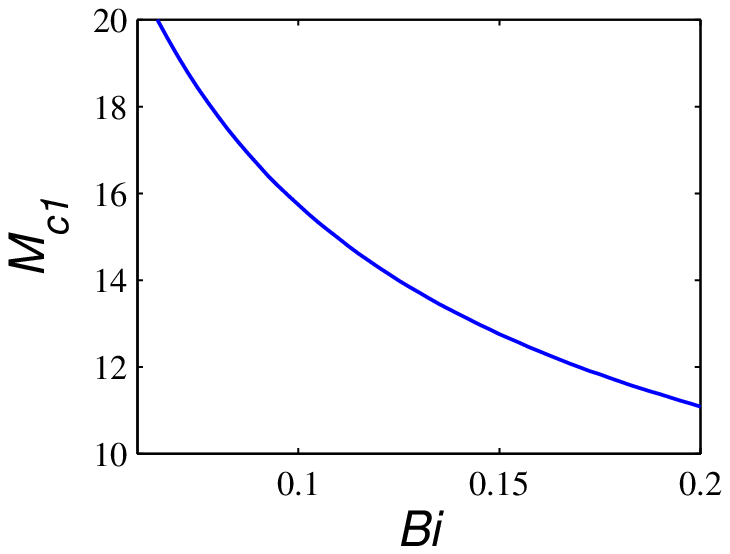}
		\caption{Marginal stability curve for oscillatory instability.}
		\label{fig:Mcrit_Bi}
	\end{minipage}
\end{figure}
For the non-volatile film, the domain $-80 \leq X \leq 80$ of length $2L = 160$ is divided into a uniform grid of 1001 points. A finite difference scheme based on \citet{fornberg1988} is used to discretize equation \ref{lh_evolutioneqn}. MATLAB function fsolve is used to solve the resulting set of equations for base state ($H_0(X)$) corresponding to given values of parameters $M$ and \Bi, with the boundary conditions replaced by \ref{vf_evolutioneqn_bcs}. For a volatile film, similar scheme is used to compute the base state from equation \ref{lh_evolutioneqn_volatile}.  Figure \ref{fig:baseprofiles} shows base profiles $H_0(X)$ for non-volatile film with two sets of values of parameters $M$ and \Bi, along with a base state profile for volatile film.  It can be seen that for both types of films, the base state film profile is flat far upstream and downstream, while a capillary ridge exists at the leading edge of the heater. Though the height of capillary ridge changes with parameters $M$ and $Bi$ for a non-volatile film and with parameters $K$, $E$ and $M$ for volatile film, the shape of the base state film profile remains qualitatively same. Note that some mass is lost due to the error in finite difference approximation near the capillary ridge. Hence, the film thickness downstream, $H_0(X=L)$ is slightly less than the upstream film thickness $H_0(X=-L)=1$, which is otherwise not expected for a non-volatile film.

Linearizing equation \ref{lh_evolutioneqn} for a small disturbance gives (see Appendix \ref{appA})
\begin{equation}
\frac{d H}{d t} = A H
\label{eq_linstab_nvol}
\end{equation}
The leading eigenvalue and corresponding eigen vector of matrix $A$ that govern the linear stability of the film are computed using MATLAB function eig.  The base state is then perturbed using the eigen vector corresponding to the eigen value with largest real part.
\begin{figure}
\centering
	\begin{subfigure}[t]{0.48\textwidth}
	\centering
		\includegraphics[width=\textwidth]{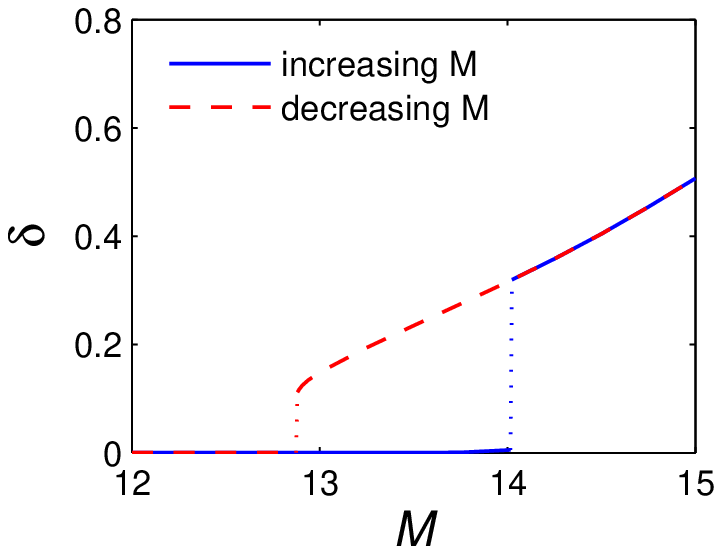}
		\caption{}
		\label{fig:deltarev_Bip125}
	\end{subfigure}
	\quad
	\begin{subfigure}[t]{0.48\textwidth}
	\centering
		\includegraphics[width=\textwidth]{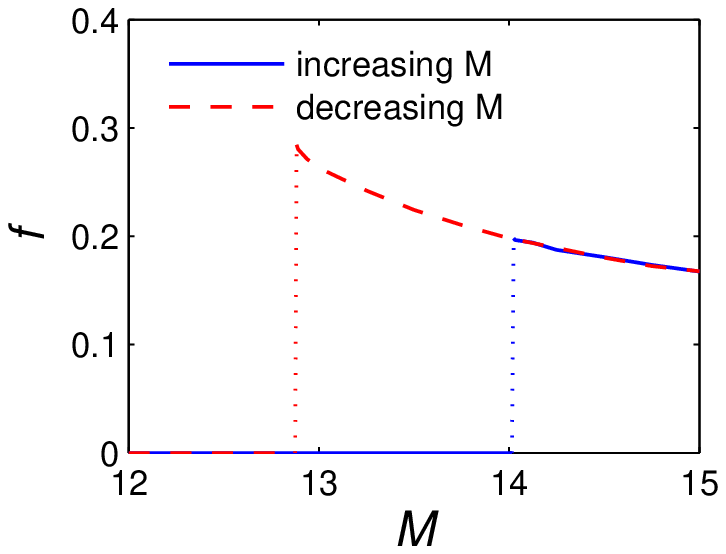}
		\caption{}
		\label{fig:freqrev_Bip125}
	\end{subfigure}
	\caption{(a) Amplitude and (b) frequency of oscillations as a function of $M$ of a non-volatile film for $Bi=0.125$. A hysteresis region is seen for $M_{c2} < M < M_{c1}$.}
\end{figure}
For a constant value of \Bi, $M$ is gradually increased and the eigenvalues and eigen vector are computed for each $M$.  For smaller $M$, the base state of the film $H_0(X)$ is stable and hence the leading eigenvalue is real and negative.  However, at a critical value $M_{c1}$, the leading eigenvalue becomes complex with the real part close to zero. Bisection method is used to compute $M_{c1}$, within an accuracy of $10^{-2}$, for different values of $Bi$ (Figure \ref{fig:Mcrit_Bi}). It is observed that with increase in \Bi, $M_{c1}$ decreases. Dynamics of the perturbed state are studied by integrating discretized form of equation \ref{lh_evolutioneqn} in time using the perturbed state as initial condition.  The film either returns back to the base state (for a stable base state) or reaches a new state with stable oscillations (for an unstable base state).  Integration is performed using the method of lines algorithm implemented in MATLAB solver ode15s and the value of maximum film height $H_{max}(t)$ is noted.  For a stable base state, $H_{max}(t)$ is nearly constant over time. On the other hand, for an unstable base state, $H_{max}(t)$ oscillates with time.  The amplitude of oscillations $\delta$ and its frequency $f$ are measured from the data.  Time integration is continued until $\delta$ reaches a constant value within a tolerance of $10^{-4}$ and $f$ reaches a constant value within a tolerance of $10^{-2}$. 

\begin{figure}
	\begin{subfigure}[t]{0.48\textwidth}
		\includegraphics[width=\textwidth]{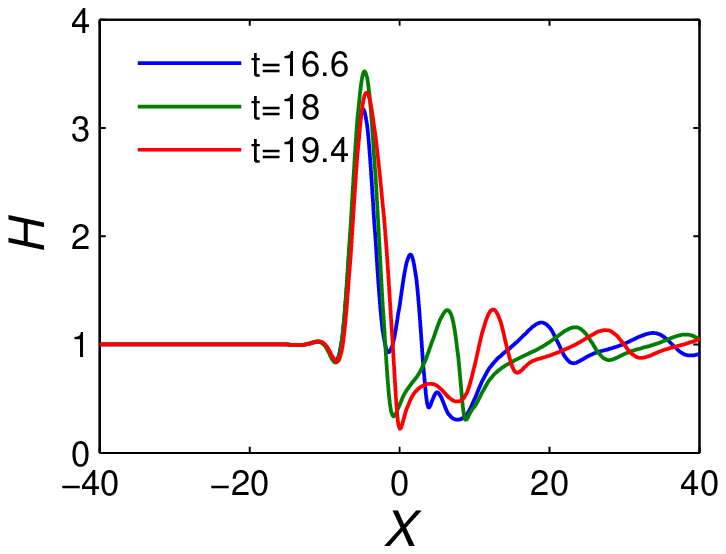}
		\caption{$M = 14 > M_{c1}$.}
		\label{fig:oscillations_14}
	\end{subfigure}
	\quad
	\begin{subfigure}[t]{0.48\textwidth}	
		\includegraphics[width=\textwidth]{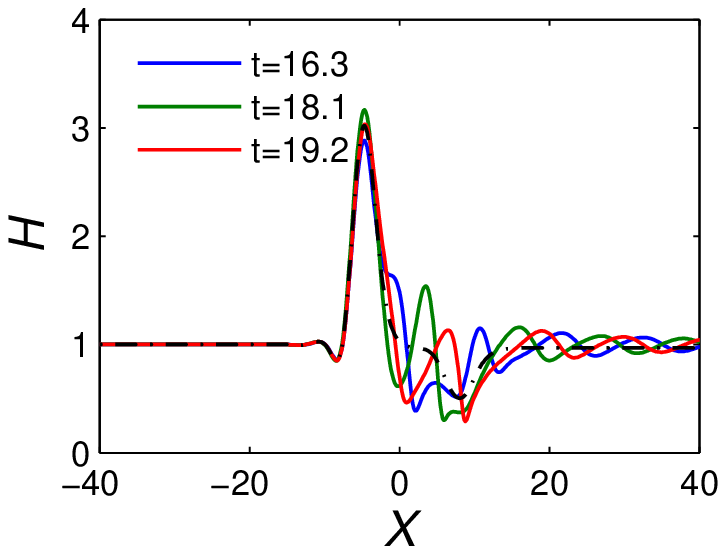}
		\caption{$M_{c2}<M = 13<M_{c1}$ (Hysteresis region)}
		\label{fig:oscillations_13}
	\end{subfigure}
	\caption{Oscillations in non-volatile film profile at different times, for  $Bi = 0.125$.  Dashed line in (b) is the base steady state. Plots are truncated at $x = \pm 40$ for better visibility.}
\end{figure}

For a typical $50 \mu$m thin film of water at ambient conditions, assuming $\gamma_T = 1.8$ dyne/cm$^\circ$C \citep{Burelbach1988}, Marangoni number $M$ is in the range of $10-15$ for temperature increase of the order of $20-30 ^\circ$C at the heater. Of the two parameters that govern the system stability, $Bi$ is kept at a constant value of $0.125$ and $M$ is gradually increased up to $M=15$. For every value of $M$, the base state $H_0(X)$ is obtained by solving \ref{lh_evolutioneqn}, using the base state of slightly smaller $M$ as initial guess. $\delta$ and $f$ are then computed as described. A plot of $\delta$ as a function of increasing $M$ for $Bi=0.125$ is shown as a solid blue line in figure \ref{fig:deltarev_Bip125}.  For $M<M_{c1}$ where $M_{c1} \simeq 13.99$, $\delta$ remains close to $0$ within tolerances. Thus, the base film profile $H_0(X)$ is stable for small perturbations.  For $M>M_{c1}$, however, the base profile becomes unstable and the film reaches an oscillatory stable state, characterized by a non-zero value of $\delta$. With further increase in $M$, the amplitude of oscillations keeps on increasing. Oscillations in film profile at $M=14$ ($>M_{c1}$) for different times are shown in figure \ref{fig:oscillations_14} . The transition at $M=M_{c1}$ can also be seen from figure \ref{fig:freqrev_Bip125} that shows the frequency of oscillations $f$ as a function of $M$ for $Bi=0.125$. The frequency of oscillations is calculated only if $\delta>10^{-2}$.
Starting with the oscillatory stable state for $M=15$, $M$ is gradually decreased in order to study hysteresis behavior of the system. For each value of $M$, the oscillatory solution for slightly higher value of $M$ is taken as initial condition to integrate equation \ref{lh_evolutioneqn} in time to obtain $\delta$ and $f$. The dashed red curve in figure \ref{fig:deltarev_Bip125} shows a plot of $\delta$ versus $M$ for decreasing $M$. It can be seen that for $M>M_{c2}$ ($M_{c2} \simeq 12.90$), the oscillatory state is stable. Thus, an initially stable oscillating film will continue to oscillate even for $M<M_{c1}$.  In the region between $M_{c2}$ and $M_{c1}$, both the base state and the oscillatory state are stable.  Figure \ref{fig:oscillations_13} shows snapshots of film profile for $M=13$ at different times, after the film reaches steady oscillations. Supplementary figures (online) show similar behavior for $Bi=0.07$.  The values of $M_{c2}$ and $M_{c1}$ are approximately $18.29$ and $19.26$ respectively.
\subsection*{Volatile film}
For a volatile film, linearized equation similar to equation \ref{eq_linstab_nvol} is obtained for growth rate of a small perturbation (see Appendix \ref{appA}). The base state is perturbed using the fastest growing eigen vector obtained from the linearized equation and used as an initial condition to integrate equation \ref{lh_evolutioneqn_volatile} in time. MATLAB function ode15s is used to perform the time integration, which is carried out until a constant value of $\delta$ within a tolerance of $10^{-4}$ is obtained.
\begin{figure}
\centering
	\begin{subfigure}[t]{0.48\textwidth}
		\includegraphics[width=\textwidth]{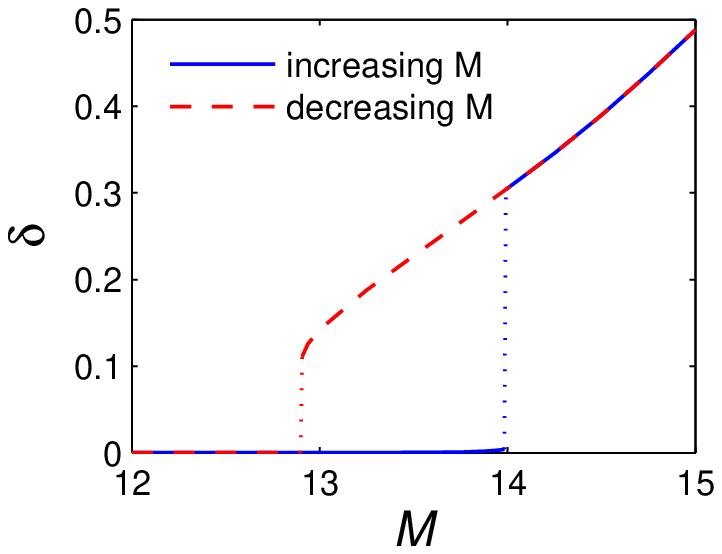}
		\caption{}
		\label{fig:vf_deltarev_Em8}
	\end{subfigure}
	\quad
	\begin{subfigure}[t]{0.48\textwidth}
		\includegraphics[width=\textwidth]{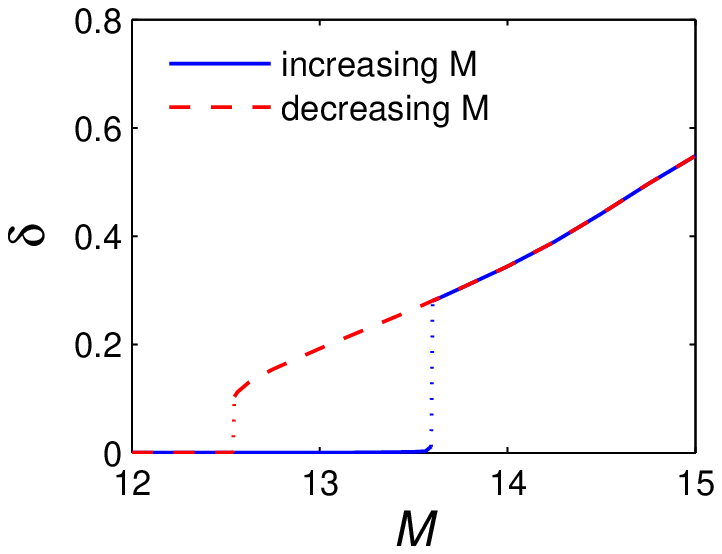}
		\caption{}
		\label{fig:vf_deltarev_Em1}
	\end{subfigure}
	\caption{Amplitude of oscillations of a volatile film as a function of $M$ for (a) $E=10^{-8}$ and (b) $E=10^{-1}$, for $K=8$ showing a hysteresis region for $M_{c2} < M < M_{c1}$.}
\end{figure}

The parameter $K$ is set constant to $8$ to facilitate direct comparison with the results for a non-volatile film with $Bi=0.125$ (see \citet{Tiwari2009} for choice of parameter values).  For a given $E$, $M$ is gradually increased as described earlier, to get a plot of $\delta$ versus $M$ for increasing $M$.  Beyond $M=M_{c1}$, the base state is no longer stable.  Instead, the film reaches a stable oscillating state characterized by a finite $\delta$.  Another plot of $\delta$ versus $M$ is obtained for decreasing $M$, where time integration is performed using the stable oscillating solution for slightly higher $M$ as initial condition.

Plots of $\delta$ vs $M$ for two different values of $E$ for a volatile film are shown in figures \ref{fig:vf_deltarev_Em8} and \ref{fig:vf_deltarev_Em1}. The solid blue line indicates results for increasing $M$, while the dashed red curve is for decreasing $M$.  Similar to the non-volatile films, these curves do not overlap completely for the range of $E$ studied.

\section{Discussion and conclusion}\label{sec:discussion}

In a typical Hopf bifurcation (see \citet{strogatz1994}), a conjugate pair of complex eigenvalues crosses the origin as the controlling parameter increases, resulting into a positive real part.  Thus, a stable fixed point becomes unstable at a critical parameter value.  If a stable limit cycle with small amplitude is formed surrounding the unstable fixed point, the system changes from an equilibrium fixed point solution to a stable limit cycle.

For a subcritical Hopf bifurcation, two stable states (a steady state and an oscillatory state) exist prior to the parameter ($M$ in this work) crossing its critical value.  A stable fixed point is surrounded by an unstable inner limit cycle of smaller amplitude, which is further enclosed by larger stable limit cycle around it.  Thus, for small amplitude perturbations from the fixed point, the system returns back to the fixed point. As the parameter changes towards its critical value ($M_{c1}$), the inner limit cycle shrinks. At the critical value of parameter, the inner limit cycle overlaps the fixed point, making it unstable. Thus, beyond critical parameter value, the system moves to the only stable outer limit cycle. The amplitude of this outer limit cycle typically depends on the parameter value.  But even at the critical parameter value, the outer limit cycle has a finite non-zero amplitude.  Now if the parameter value changes in the other direction, after crossing the critical value, the fixed point becomes stable. However, the system stays on the outer limit cycle, as small perturbations are not enough to drive the system back to the fixed point. This continues until the outer limit cycle shrinks to the inner limit cycle and becomes unstable ($M_{c2}$). Thus, a hysteresis is observed in a typical subcritical Hopf bifurcation.

Bifurcation diagram for non-volatile film for $Bi=0.125$ (figures \ref{fig:deltarev_Bip125} and \ref{fig:freqrev_Bip125}) shows that for $M<13.99$, a stable fixed point solution exists.  In other words, the base state solution similar to the one shown in figure \ref{fig:schematic} is stable. A small disturbance in the base film profile will decay with time and the film will return to the base state. As $M$ increases further, beyond a particular value $M_{c1} = 13.99$, the base state becomes unstable.  A small perturbation in the base film profile grows in time.  The growth of perturbation continues until another stable state is reached.  The new stable state that the film reaches is an oscillatory state.  As seen from the figures, the oscillatory state has a finite non-zero amplitude $\delta = 0.30$ and frequency of oscillations $0.20$ even for $M$ slightly greater than $M_{c1}$.  With further increase in $M$, the amplitude of oscillations keeps on increasing while the frequency decreases.

From linear stability analysis, the critical value of $M$ at which the eigenvalues with largest real part cross the origin and become positive for $Bi = 0.125$ is $M_{c1} = 13.99$ (figure \ref{fig:Mcrit_Bi}), which is in good agreement to $M_{c1}=13.99$ at which the bifurcation takes place. Thus, linear stability theory correctly predicts the critical point within acceptable tolerances.

Starting with the oscillatory state, if $M$ is gradually decreased, the film continues to exhibit oscillatory state with decreasing amplitude. However, even when $M<M_{c1}$ the film keeps on oscillating instead of returning to the stable base state.  This continues until $M = 12.90$, when the film returns from oscillations with amplitude $\delta = 0.11$ and frequency $f = 0.29$ to the base state ($\delta = 0$).  Thus, at $M$ slightly greater than $M_{c1}$, the stable fixed point becomes unstable and the system jumps to a finite amplitude limit cycle. Also, starting on a limit cycle, even when $M$ decreases beyond $M_{c1}$, the system stays on the limit cycle and returns to the fixed point only at a much smaller value of $M = 12.90$.  This indicates that the film exhibits a hysteresis region and undergoes a subcritical Hopf bifurcation at $M_{c1} = 13.99$.

The nature of bifurcation remains qualitatively similar for $Bi = 0.07$ (Supplementary figures).  A subcritical Hopf bifurcation is seen at $M_{c1} = 19.26$ where the film jumps from base state ($\delta = 0$) to an oscillatory state with amplitude $\delta = 0.30$ and frequency $f = 0.21$. Hysteresis is observed while continuing for decreasing $M$, with the system returning to the base state only at $M = 18.29$ where the stable finite amplitude oscillations $\delta = 0.14$ and frequency $f = 0.27$ disappear.

Using the volatile film equations and setting $E = 0$ should recover results for the non-volatile film.  An agreement is found between the bifurcation diagrams (not shown) of volatile film with $E=0$, $K=8$ and that of a non-volatile film with $Bi = 0.125$. The bifurcation diagram for volatile film with $K = 8$ and $E = 10^{-8}$ is shown in figure \ref{fig:vf_deltarev_Em8}. Note that $E = 10^{-8}$ corresponds to a weakly evaporative film. It can be seen that even a volatile film shows a subcritical bifurcation similar to that observed for a non-volatile film, with a hysteresis region between $M = 13.99$ and $M=12.90$.  For a volatile film with $K = 8$ and $E = 10^{-1}$, the hysteresis region lies between $M = 13.60$ and $M = 12.55$.

In conclusion, we were able to analyze the stability of a thin film flowing over a localized heater using an evolution equation for a two-dimensional film. An oscillatory instability is observed beyond a critical point in the parameter space (Marangoni number).  The onset of instability predicted by the linear stability theory agrees with the observation from the numerical non-linear film dynamics. While continuing along Marangoni number, we observe a hysteresis region where two stable states exist at the same values of parameters. Depending on its initial state, the film either continues to be at a steady base state as shown in figure \ref{fig:baseprofiles} or exhibits steady oscillations as shown in figure \ref{fig:oscillations_13}. The presence of two stable solutions in the hysteresis region is well captured by the simplified evolution equation that governs the balance of different forces in the film. The hysteresis region is present in both non-volatile and volatile films and is identified as a sub-critical Hopf bifurcation.  The presence of this bifurcation at critical parameter values, that are within the typically explored range in systems involving thin liquid films flowing down heated surfaces, can have profound effects on the design and operation of systems such as falling film evaporators.
\appendix
\section{}\label{appA}

For a non-volatile film, a small perturbation applied to the base state can be represented as $H_1(X,t) = H_0(X) + \epsilon H(X,t)$, where $\epsilon \ll 1$. Substituting in equation \ref{lh_evolutioneqn}, the leading order linearized evolution equation that governs the growth of perturbation is
\begin{equation}
\begin{array}{rcl}
H_t &=& a_0 H + a_1 H_X + a_2 H_{XX} + a_3 H_{XXX} + a_4 H_{XXXX}
\label{lh_stabilityeqn}
\end{array}
\end{equation}
with the coefficients $a_i$ given by
\begin{equation}
\begin{array}{rcl}
a_0 &=& 3 H_0^2 H_{0_{XXXX}} + 6 H_0 H_{0_X} (1+H_{0_{XXX}}) + 2 M H_0 H_{0_X} T^i_{1_{X}} + M H_0^2T^i_{1_{XX}}\\
    & &- 2 M (H_{0_X} T^i_{0_X} + H_0 T^i_{0_{XX}})\\
a_1 &=& 3 H_0^2 (1+H_{0_{XXX}}) + M H_0^2 T^i_{1_{X}} + 2 M H_0 H_{0_{X}} T^i_1 + M H_0^2 T^i_{1_{X}} - 2 M H_0 T^i_{0_{X}}\\
a_2 &=& M H_0^2 T^i_1\\
a_3 &=& 3 H_0^2 H_{0_{X}}\\
a_4 &=& H_0^3
\end{array}
\end{equation}
with $T^i_1 = \theta_0 Bi /(1 + Bi H_0)^2$ and $T^i_0 = \theta_0 / (1 + Bi H_0)$.  The subscripts $_t$ and $_X$ indicate a derivative with respect to $t$ and $X$ respectively. Using finite difference approximation for derivatives of $H$ results into equation \ref{eq_linstab_nvol}.

Similarly, for a volatile film,  substituting $H_1(X,t) = H_0(X) + \epsilon H(X,t)$ in the governing equation \ref{lh_evolutioneqn_volatile} gives the growth of small perturbation $H(X,t)$ similar to equation \ref{lh_stabilityeqn}, with the coefficients given by
\begin{equation}
\begin{array}{rcl}
a_0 &=& -(E \theta_0 /(K+H_0)^2 - 6 H_0 H_{0_{X}} (1+H_{0_{XXX}}) - 3 H_0^2 H_{0_{XXXX}}) + 2 M H_0 H_{0_{X}} T^i_{1_{X}} \\
& & + 2 M H_{0_{X}} T^i_{0_{X}}  + M H_0^2 T^i_{1_{XX}} + 2 M H_0 T^i_{0_{XX}})\\
a_1 &=& -(-3 H_0^2 (1+H_{0_{XXX}}) + 2 M H_0 H_{0_{X}} T^i_1  + 2 M H_0 T^i_{0_{X}}) + 2 M H_0^2 T^i_{1_{x}}\\
a_2 &=& -(M H_0^2 T^i_1)\\
a_3 &=& -(-3 H_0^2 H_{0_{X}})\\
a_4 &=& -(-H_0^3)
\end{array}
\end{equation}
with $T^i_0 = K \theta_0 /(K+H_0)$ and $T^i_1 = -K \theta_0/(K+H_0)^2$.
\bibliographystyle{jfm}
\bibliography{paper1}
\end{document}